\documentclass[prd,twocolumn,a4paper,superscriptaddress,floatfix]{revtex4}
\usepackage{graphicx}
\usepackage{bbm}
\begin{document}

\newcommand{\be}{\begin{equation}}
\newcommand{\ee}{\end{equation}}
\newcommand{\bq}{\begin{eqnarray}}
\newcommand{\eq}{\end{eqnarray}}
\newcommand{\bsq}{\begin{subequations}}
\newcommand{\esq}{\end{subequations}}
\newcommand{\bc}{\begin{center}}
\newcommand{\ec}{\end{center}}
\newcommand\lsim{\mathrel{\rlap{\lower4pt\hbox{\hskip1pt$\sim$}}
    \raise1pt\hbox{$<$}}}
\newcommand\gsim{\mathrel{\rlap{\lower4pt\hbox{\hskip1pt$\sim$}}
    \raise1pt\hbox{$>$}}}
\newcommand\esim{\mathrel{\rlap{\raise2pt\hbox{\hskip0pt$\sim$}}
    \lower1pt\hbox{$-$}}}

\title{$p$-brane dynamics in $N+1$-dimensional FRW universes}

\author{P.P. Avelino}
\email[Electronic address: ]{ppavelin@fc.up.pt}
\affiliation{Centro de F\'{\i}sica do Porto, Rua do Campo Alegre 687, 4169-007 Porto, Portugal}
\affiliation{Departamento de F\'{\i}sica da Faculdade de Ci\^encias
da Universidade do Porto, Rua do Campo Alegre 687, 4169-007 Porto, Portugal}
\author{R. Menezes}
\email[Electronic address: ]{rms@fisica.ufpb.br}
\affiliation{Centro de F\'{\i}sica do Porto, Rua do Campo Alegre 687, 4169-007 Porto, Portugal}
\affiliation{Departamento de F\'{\i}sica da Faculdade de Ci\^encias
da Universidade do Porto, Rua do Campo Alegre 687, 4169-007 Porto, Portugal}
\affiliation{Departamento de F\'{\i}sica, Universidade Federal da Para\'{\i}ba
58051-970 Jo\~ao Pessoa, Para\'{\i}ba, Brasil}
\author{L. Sousa}
\email[Electronic address: ]{laragsousa@gmail.com}
\affiliation{Centro de F\'{\i}sica do Porto, Rua do Campo Alegre 687, 4169-007 Porto, Portugal}
\affiliation{Departamento de F\'{\i}sica da Faculdade de Ci\^encias
da Universidade do Porto, Rua do Campo Alegre 687, 4169-007 Porto, Portugal}

\date{10 November 2008}
\begin{abstract}
We study the evolution of maximally symmetric $p$-branes with a $S_{p-i}\otimes \mathbbm{R}^i$ topology in flat expanding or collapsing homogeneous and isotropic universes with $N+1$ dimensions (with $N \ge 3$, $p < N$, $0 \le i < p$). We find the corresponding equations of motion and compute new analytical solutions for the trajectories in phase space. For a constant Hubble parameter, $H$, and $i=0$ we show that all initially static solutions with a physical radius below a certain critical value, $r_c^0$, are periodic while those with a larger initial radius become frozen in comoving coordinates at late times. We find a stationary solution  with constant velocity and physical radius, $r_c$, and compute the root mean square velocity of the periodic $p$-brane solutions and the corresponding (average) equation of state of the $p$-brane gas. We also investigate the $p$-brane dynamics for $H \neq {\rm constant}$ in models where the evolution of the universe is driven by a perfect fluid with constant equation of state parameter, $w={\cal P}_p/\rho_p$, and show that a critical radius, $r_c$, can still be defined for $ -1 \le w < w_c$ with 
$w_c=(2-N)/N$. We further show that for $w \sim w_c$ the critical radius is given approximately by $r_c H \propto (w_c-w)^{\gamma_c}$ with $\gamma_c=-1/2$ ($r_c H \to \infty$ when $w \to w_c$). Finally, we discuss the impact that the large scale dynamics of the universe can have on the macroscopic evolution of very small loops.

\end{abstract}
\maketitle

\section{\label{intr}Introduction}
The formation of topological defect networks is expected at cosmological phase transitions in the early universe \cite{Kibble,Book}. 
The cosmological consequences of such phase transitions can be very diverse depending both on the type of defects formed and on the evolution of the universe after the phase transition. For example, it is well known that the average domain wall density in an expanding universe grows with respect to the background density in the radiation and matter eras and 
consequently domain walls must necessarily be very light in order not to completely dominate the energy density of the universe 
\cite{ZEL}. On the other hand, this late time dominance is a property naturally associated with the dark energy and domain wall networks have been proposed as interesting dark energy candidates \cite{Bucher:1998mh}. However, a domain wall network would need to have a very small characteristic length and velocity in order to be the dark energy. This requirement has recently been shown not to be fulfilled both in the simplest domain wall models and in more complex scenarios with junctions \cite{PinaAvelino:2006ia,Avelino:2006xf}.

There are other topological defects, such as cosmic strings, which lead to much less dramatic consequences. A cosmic string network, in the scaling regime, has an average density that is roughly proportional to the background density. Consequently standard cosmic strings do not tend to dominate the energy density of the universe and they naturally generate a scale-invariant spectrum of density perturbations on cosmological scales. Although recent cosmological data, in particular the cosmic microwave background anisotropies, rules out cosmic strings as the main source of density perturbations on large cosmological scales, a cosmic string contribution at a subdominant level is all but excluded \cite{Durrer:2001cg,Pogosian:2003mz}. In fact, the interest in cosmic strings has recently been revived by fundamental work suggesting string production at a significant (but subdominant) level at the end of brane inflation \cite{Sarangi:2002yt}. 

Cosmic strings and other defects can also be formed during an inflationary era or in between several periods of inflation. It is thus crucial to understand their evolution in such regimes in order to quantify their ability to survive any inflationary period which might occur after they form \cite{basu,openinf}. The evolution of circular cosmic string loops and spherical domain walls in a flat FRW universe has previously been studied in  \cite{loopinf} (see also \cite{Letelier:1990qw}) where the existence of periodic solutions in a de Sitter universe has been demonstrated. Here, we generalize these results in particular by explicitly computing the phase space trajectories and determining the critical radius for spherical $p$-branes in space-times with an arbitrary number of dimensions. We also study in detail the more general evolution of maximally symmetric $p$-branes with a $S_{p-i}\otimes \mathbbm{R}^i$ topology in expanding and collapsing flat Friedmann-Robertson-Walker  (FRW) universes. 

The outline of this paper is as follows. In section II we derive the equations of motion for spherically symmetric $p$-branes in 
FRW universes with an arbitrary number of spatial dimensions. We generalize these results in Section III to account for maximally symmetric $p$-branes with a $S_{p-i}\otimes \mathbbm{R}^i$ topology. In section IV we study in detail 
the dynamics of $p$-branes in expanding and collapsing universes with a constant Hubble parameter, $H$, computing, 
in particular, the phase space trajectories and the critical radius, $r_c$, for spherically symmetric $p$-branes and discussing 
$p$-brane evolution in various limits. In section V we relax the constant $H$ condition and consider the possibility of a universe dominated by a dense $p$-brane gas. We determine sufficient conditions for the cosmology to have an impact on the macroscopic dynamics of small spherically symmetric $p$-branes in section VI and we conclude in section VII.
 
\section{\label{sph} Spherical symmetric $p$-branes}

The space-time trajectory of a $p$-brane in a $N$+1-dimensional space-time is described by the 
Nambu-Goto action
\be
S_p=-\tau_p\int d^{p+1} \zeta {\sqrt {-\gamma}}\,,
\ee
where $\tau_p$ is the $p$-brane tension, $\gamma_{ab}=g_{\mu \nu}x_{,a}^{\mu}x_{,b}^{\nu}$, $g_{\mu\nu}$ are the metric components and $x^\mu=x^\mu(\zeta^a)$  represents the space-time trajectory of the $p$-brane ($a,b=0,\dots,p$). 

In a $N$+1-dimensional Minkowski spacetime, the trajectory of a $p$-brane with spherical symmetry may be written, in hyperspherical coordinates, as
\bq
x(t,\phi_1,\dots,\phi_{p-1})=q(t)\left[\cos\phi_1,\sin\phi_1\cos\phi_2,\right.\nonumber\\
\left.  \sin\phi_1 \sin\phi_2\cos\phi_3,\dots,\right.\nonumber\\
\left.  \sin\phi_1 \dots\sin\phi_{p-2}\cos\phi_{p-1},\right.\nonumber\\
\left.  \sin\phi_1 \dots \sin\phi_{p-1},0,\dots,0\right]\,,
\eq
where $\phi_1,\dots,\phi_{p-2}\in \left[0,\pi\right[$ and $\phi_{p-1}\in \left[0,2\pi\right[$ and for simplicity the coordinate system was chosen in order to make the defect aligned with the first $p$-dimensions. The area of a $p$-dimensional spherically symmetric $p$-brane is the area of a $p$-dimensional hypersphere,
\be
S_p=(p+1)C_{p+1}|q|^{p}\,,
\ee
where $|q|$ is the physical radius and 
\be
C_j=\frac{\pi^{j/2}}{\Gamma\left(j/2+1\right)}.
\ee
The energy of the $p$-brane is, then, proportional to $S_p\gamma$, where $v=dq/dt$ and $\gamma=(1-v^2)^{-1/2}$. It then 
follows from energy conservation that
\be
\frac{dR}{dt}=0\,,
\frac{dv}{dt}=(1-v^2)\left(\frac{p \gamma^{1/p} s}{R}\right)\,,
\label{micvloop}
\ee
where $R=|q|\gamma^{1/p}$ is the \textit{invariant} radius and $s=sign{(-q)}$.

In a flat $N$+1-dimensional Friedmann-Robertson-Walker Universe the energy of the $p$-brane is no longer conserved and, consequently, the value of $R$ is expected to depend on time. Energy-momentum conservation for a planar $p$-brane implies that the momentum per comoving area (that is, the $p$-dimensional generalization in a $N$+1-dimensional spacetime of the concept of the usual 2-dimensional area in a 3+1-dimensional spacetime) is proportional to $a^{-1}$ and consequently the velocity of a planar $p$-brane satisfies $v\gamma\propto a^{-(p+1)}$ which implies
\be
\frac{dv}{dt}=-(1-v^2)(p+1)vH\,.
\ee
The equation of motion for a spherically symmetric $p$-brane has an additional curvature term (see Eqn. (\ref{micvloop})) and  is 
then given by
\be
\frac{dv}{dt}=(1-v^2)\left[\frac{p \gamma^{1/p}s}{R}-(p+1)vH\right]\,,
\ee
so that
\be
\frac{dR}{dt}=HR\left[1-\frac{p+1}{p}v^2\right]\,,
\ee
where $R$ is now defined as $R=\gamma^{1/p} |r|$ with $r=a q$. Note that, in this case, if  $H \neq 0$ then $R$ is no longer 
time-independent.

\section{\label{sph} $S_{p-1}\otimes \mathbbm{R}$ $p$-branes}

Let us generalize the results of the previous section for maximally symmetric $p$-branes with a $S_{p-i}\otimes \mathbbm{R}^i$ topology, 
with $p < N$ and $0 \le i \le p$, by allowing a number $i$ of dimensions to have no curvature and leaving $p-i$ dimensions with spherical symmetry. In this case the area of the defect (per unit of i-dimensional area of the non-curved dimensions) is
\be
S_p^i=(p-i+1)C_{p-i+1}\gamma |q|^{p-i}\,
\ee
and energy-momentum conservation leads to the following equations of motion
\bq
\frac{dv}{dt} &=& (1-v^2)\left[\frac{(p-i)\gamma^{1/(p-i)}s}{R}-(p+1)vH\right] \,,\label{micvloop1a}\\
\frac{dR}{dt} &=& HR\left[1-\frac{p+1}{p-i\, }v^2\right]\,, \label{micvloop1b}
\eq
where the so-called invariant radius, $R$, is now given by $R=|q|\gamma^{1/(p-i)}a$. 
Note that 
\bq
\frac{dr}{dt} &=& v+Hr\,,\label{lph1}\\
v &=& \pm \sqrt{1-\left(\frac{r}{R}\right)^{2(p-i)}}\,. \label{lph2}
\eq
These equations clearly show that there is an asymmetry between the collapse and the expansion of the p-brane. 
This asymmetry is clearly seen in Fig. 1 which is discussed in detail in the following section. Note that If the universe is 
expanding then there is a damping term which always contributes to decrease $|v|$. On the other hand, the curvature term may act 
to increase or decrease $|v|$ depending on whether the p-brane is collapsing or expanding. It is also very interesting to realize that we may write the Hubble parameter as a function of the 
two loop parameters, $R$ and $v$, as
 \be
H=\frac{d \ln R}{dt}\left(1-\frac{p+1}{p-i} v^2\right)^{-1},
\ee
What this equation clearly shows is that we could, in principle, infer the dynamics of the universe by looking at the dynamics of a 
single $p$-brane.

Note that the equations of motion 
(\ref{micvloop1a}) and  (\ref{micvloop1b}) are invariant with respect to the transformation $q \to -q$ and 
$t \to -t$. This implies that the phase space trajectories in expanding and collapsing universes 
related by the transformation $H \to -H$ are identical. However, the transformation $t \to -t$ implies that 
the direction in which the trajectories are travelled is reversed.

\section{p-brane dynamics (H=constant)}

We have determined the $(r,v)$ trajectories in phase space in the case of spherically symmetric $p$-branes ($i=0$). Integrating 
Eqns. (\ref{micvloop1a}) and (\ref{micvloop1b}) we have
\be
\gamma r^p(1+vHr)=k\,,
\ee
where $\gamma=(1-v^2)^{-1/2}$ and $k$ is a constant. A stationary solution of Eqns. (\ref{micvloop1a}) and (\ref{micvloop1b}) with fixed critical velocity ($\dot v=0$) and physical radius ($\dot r=0$), respectively $v_c$ and $r_c$, is characterized by
\be
v_c^2  =  H^2 r_c^2 = \frac{p-i}{p+1}\,. \label{statphys}
\ee
This solution describes a $p$-brane standing still against the Hubble expansion  or contraction. If $i=0$ then 
$v_c^2= p/(p+1)$ and we may easily find that value of $k$ corresponding to the stationary solution is given by $k=k_c=|H|^{-p} p^{p/2}(p+1)^{-(p+1)/2}$. Trajectories with $k > k_c$ will approach the line defined by $v=-(Hr)^{-1}$ when $|r| \to \infty$ (so that 
$v \to 0$ in this limit). On the other hand all periodic trajectories have $k<k_c$. Note that the reverse is not true since there is an infinite number of non-periodic trajectories with $k<k_c$. This is clearly shown in Fig 1 where spherical domain walls trajectories with $p=2$ and $i=0$ 
are represented in phase space. The trajectories with $k<k_c$ are represented by a solid line and the dashed lines represent trajectories with $k>k_c$ which start at the one of the critical points defined by  $v=-Hr=\pm 1$ and end at $(r=\infty,v=0)$ in an expanding universe (or vice-versa in a collapsing universe). Note that even the periodic trajectories, for $k<k_c$, are asymmetric. This is a direct result of the asymmetry of the equations of motion (\ref{lph1}) and  (\ref{lph2}) discussed in Section III.

\begin{figure}[ht!]
\vspace{1cm}
\includegraphics[width=8cm,angle=0]{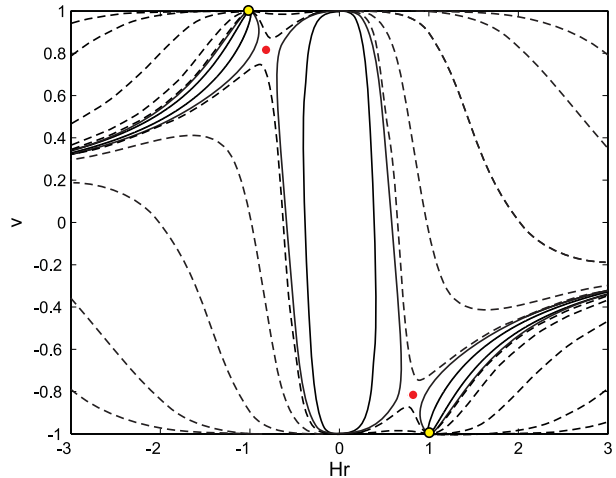}
\caption{\small{Representation of the trajectories of spherical domain walls ($p=2$, $i=0$) in phase space. The critical points marked with a red dot represent the critical stationary solution defined in Eqn. \ref {statphys} and the yellow dots correspond to the critical points defined by $v=-Hr=\pm 1$. The solid lines correspond to trajectories of spherical domain walls with $k<k_c$ (those 
with $|r| < r_c$ describe periodic trajectories in phase space). The dashed lines correspond to  trajectories with $k>k_c$ which start at the one of the critical points defined by  $v=-Hr=\pm 1$ and end at $(r=\infty,v=0)$ in an expanding universe (or vice-versa in a collapsing 
universe).}}
\end{figure}

We may also compute a critical (initial) radius, $r_c^0$, corresponding to an initially static $p$-brane solution ($v^0=0$) with $k= k_c$ (in an expanding universe it will asymptote to the stationary solution when $t \to \infty$). This critical radius, is given by
\be
r_c^0=|H|^{-1} p^{1/2}(p+1)^{-(p+1)/(2p)}\,.
\ee
In the case of circular cosmic string loops ($p=1$) and spherical domain walls ($p=2$) $r_c^0=|H|^{-1}/2$ and 
$r_c^0=2^{1/2}3^{-3/4} |H|^{-1}$ respectively. We recover the critical radius, $r_c^0$, for a circular cosmic string loop 
but the critical radius, $r_c^0$, for a spherical domain wall is slightly smaller (by about $10 \%$)  than the 
approximate solution given in ref. \cite{loopinf} . For larger initially static spherically symmetric $p$-branes with 
$|r^0| > r_c^0$ the motion is not periodic and, if $H>0$, the brane eventually freezes in comoving coordinates. Specifically, 
the $p$-brane will asymptotically behave as
\be\label{largerr}
q=const. \Longleftrightarrow R\propto a\,,\quad
v\propto a^{-1}\,.
\ee
If $H<0$ then  all solutions with $|r^0|> r_c^0$ will asymptote the critical point defined by $v=-Hr=\pm 1$ when $t \to \infty$. 

\begin{figure}[ht!]
\vspace{1cm}
\includegraphics[width=8cm,angle=0]{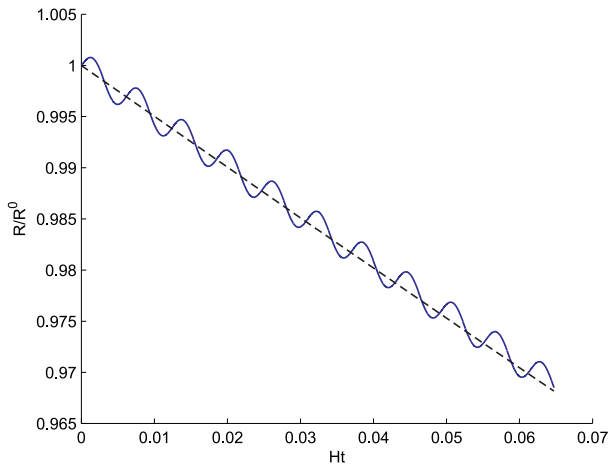}
\caption{\small{Time evolution of the invariant radius, $R$, of a domain wall with cylindrical symmetry with 
$R^0 H=0.002$ (solid line) and the predicted evolution of $\langle R \rangle$ (dashed line). \label{cyl} }}
\end{figure}

Periodic solutions with $|r^0|< r_c^0$ satisfy $\langle d(\ln R)/dt \rangle = 0$ and consequently it follows from Eqn. (\ref{micvloop1a}) with $i=0$ that
\be
\langle v^2 \rangle = \frac{p}{p+1}\label{averagevel}\,,
\ee
where the brackets denote a time average over one period. Note that the r.m.s. velocity approaches unity in the $p \to \infty$ limit. On the other hand, the energy density, $\rho_p$, in sub-critical isolated spherically symmetric $p$-branes should evolve as matter ($\rho_p \propto a^{-N})$. Since the $p$-brane equation of state is $w \equiv {\cal P}_p/\rho_p =((p+1)\langle v^2 \rangle -p)/N$  then $\langle v^2 \rangle = p/(p+1)$  \cite{Boehm,Brandenberger} is required in order that $w=0$.

If $i \neq 0$ there are no longer periodic solutions. This can easily be seen in the limit where $R |H| \ll 1$. In this limit the Hubble 
damping term has a very small impact on the dynamics of the $p$-brane on timescales $\sim R$ and consequently the dynamics of the $p$-brane is quasi periodic. Hence, from Eqn. (\ref{micvloop1a}) we see that the evolution of the velocity is essentially the same for all branes with the same value of $p-i$. In particular 
\be
\langle v^2 \rangle = \frac{p-i}{p-i+1}\label{averagevel}\,,
\ee
which only depends on $p-i$.
We may then write Eqn. (\ref{micvloop1b}) as 
\be
\frac{d R}{dt} = HR\left[1-\frac{p-i+1}{p-i} v^2 - \frac{i}{p-i} v^2\right] \label{micvloop1c}\,.
\ee
Taking the average of the r.h.s of (\ref{micvloop1a}) over one quasi period one obtains 
\bq
\frac{d {\langle R \rangle}}{dt}  &=& H \langle R \rangle \left[1-\frac{p-i+1}{p-i} \langle v^2 \rangle  - \frac{i}{p-i} \langle v^2 
\rangle \right]\,\\
&=&   - H  \langle R \rangle \frac{i}{p-i+1}\,,
\eq
so that
\be
\langle R \rangle \propto  \exp \left(\alpha_1 H t\right)\label{solmac}\,, 
\ee
with $\alpha_1=-i/(p-i+1)$ . 
Hence, if $i \neq 0$ and the universe is expanding then the $p$-brane radius shrinks over cosmological timescales (the opposite 
occurs if $H < 0$). In Fig. 2 we confirm the above results by comparing the evolution of the invariant radius, $R$, computed numerically with the analytical macroscopic solution given by Eqn.  (\ref{solmac}) and we find a good agreement between 
them.

\section{p-brane dynamics (H $\neq$ constant)}

We shall now discuss the case in which the universe is expanding and the Hubble parameter is no longer time independent. 
For simplicity we shall assume that the dynamics of the universe is driven by a fluid with $w={\rm constant} \neq -1$ so that 
$a \propto |t|^\beta$ with $\beta=2/(N(w+1))$. We identify $t=0$ either with the big-bang (for $w > -1$) or with the big rip (for $w < -1$). If $w > -1$ the Hubble radius, $H^{-1}$ increases with time but for $-1< w < w_c$ with $w_c=(2-N)/N$ (so that $\beta > 1$) the comoving Hubble radius ($H^{-1}/a$) decreases with time and consequently it is still possible to find a critical radius, $r_c$ associated with a solution that has $v \to v_c \neq 0$ when $t \to \infty$. By requiring that ${\dot v}=0$ and ${\ddot v}=0$ asymptotically at late times and using Eqn. (\ref{micvloop1a})  we find that
\bq
(r_c H) ^2 &=& \frac{p-i}{p+1} \frac{\beta}{\beta-1}\,,\\
v_c^2 &=&\frac{\beta-1}{\beta}\frac{p-i}{p+1}\,.
\eq
We see that in the $\beta \to 1$ limit $v_c \to 0$ and 
\be
r_c H \propto (\beta-1)^{\gamma_c} \propto (w_c -w)^{\gamma_c}  \to \infty\,.
\ee
The value of the critical exponent, $\gamma_c$ is equal to $-1/2$ and is independent of $i$, $p$ and $N$. 

The critical radius defined above no longer exists for $w < -1$ or $w > w_c$. If $w > w_c$ then $\beta < 1$ and consequently the comoving Hubble radius, $H^{-1}/a$, is an increasing function of time. This means that all $p$-branes (even those that are initially very large) will eventually come inside the Hubble sphere so that at late times, when $r \ll H^{-1}$, the $p$-brane will oscillate quasi-periodically. If $w < -1$ then  $\beta >1$ and the Hubble radius decreases with increasing physical time. Consequently, any initial quasi-periodically 
$p$-brane trajectory (with $r \ll H^{-1}$) will eventually eventually freeze in comoving coordinates (with $r \gg H^{-1}$).

The analysis of the collapsing case is trivial an follows directly from the  $q \to -q$ and $t \to -t$ duality described in Section III. 
In this case, if $w < w_c$ then the comoving Hubble radius, $|H|^{-1}/a$, increases with time (if $w < -1$ then $|H|^{-1}$ is also an increasing function of time). Hence, all $p$-branes  will eventually come 
inside the horizon and will oscillate quasi-periodically when $R|H|$ becomes much smaller than unity.
On the other hand, if $w > w_c$ then the comoving Hubble radius, $|H|^{-1}/a$, decreases with time and so all $p$-branes will eventually have a physical radius much larger than $|H|^{-1}$ and asymptotically the physical radius becomes $|r| \propto a$ so that $\gamma v \propto a^{-(p+1)}$. Consequently, as the universe collapses ($a \to 0$) the $p$-branes become ultra-relativistic while staying effectively frozen in comoving coordinates. This result has been demonstrated in ref. \cite{Avelino:2002xy,Avelino:2002hx} for defect networks in 3+1 dimensions but remains valid when we consider the dynamics of defect networks in contracting FRW universes with an arbitrary number of spatial dimensions as long as $w > w_c$.

An interesting case is that of a flat expanding universe dominated by a dense brane gas with average density $\rho_p$ and average pressure ${\cal P}_p$. The Einstein Equations for a $N$+1 FRW universe are given by
\bq
\frac{\ddot a}{a} &=& -\frac{8 \pi G_{N+1}}{N(N-1)} \left( (N-2) \rho_p + N {\cal P}_p \right)\,,\\
\left(\frac{\dot a}{a}\right)^2 &=& \frac{16 \pi G_{N+1}}{N(N-1)} \rho_p\,.
\eq
where $G_{N+1}$ is the $N+1$ dimensional Newton constant. If we assume a constant equation of state for the brane gas given 
by $w \equiv {\cal P}_p/\rho_p =((p+1)\langle v^2 \rangle -p)/N$ (see previous section) then the dynamics of the evolution of the 
scale factor with physical time is given by $a \propto t^\beta$ with
\be
\beta =\frac{2}{N(1+w)}= \frac{2}{N-p+(p+1)\langle v^2 \rangle}\,.
\ee
In order to accelerate the universe one needs $\beta >1$ or equivalently $w <w_c= (2-N)/N$ and consequently 
\be
\langle v^2 \rangle < \frac{2-N+p}{p+1}\,.
\ee
Here, the brackets denote an average over the brane gas network. We see that accelerated expansion is possible only if $N=p+1$ and in that case one would need $\langle v^2 \rangle < N^{-1}$ for inflation to take place. 
\begin{figure}[ht!]
\vspace{1cm}
\includegraphics[width=8cm,angle=0]{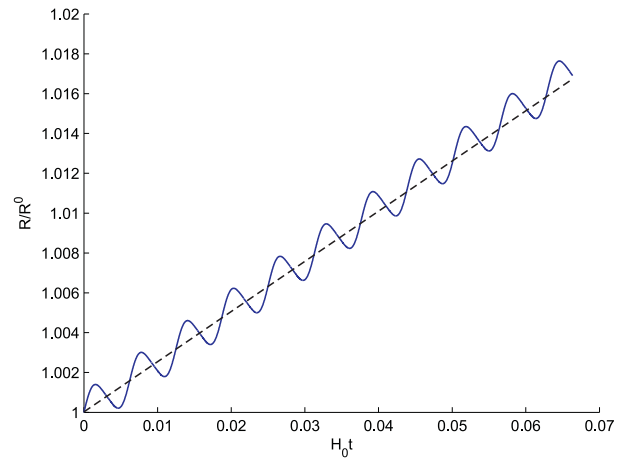}
\caption{\small{Time evolution of the invariant radius, $R$, of a circular cosmic string loop for $H(t)=H_0+H_1(1-2v^2(t))$ (solid line) and the expected evolution of the mean value of $R$ (dashed line). We assumed have taken $H_0=2H_1$ and 
$R^0 H_0 = 0.002$.  \label{loopfig} }}
\end{figure}

\section{Can Cosmology have an impact on the dynamics of small cosmic string loops ?}

In Section IV we have shown that if $i \neq 0$ then the cosmology has an impact on $p$-brane dynamics over cosmological 
timescales, even if $R|H| \ll 1$. However, we have shown that if $H={\rm constant}$ then the macroscopic evolution of sub-critical spherically symmetric $p$-branes ($i=0$) is not influenced by the cosmology ($\langle R \rangle={\rm constant}$). In this section 
we ask whether or not there are cosmological models with a Hubble parameter, $H(t)$, in which the macroscopic dynamics of 
small spherically symmetric $p$-branes can be affected by the large-scale evolution of the universe.

If we consider the evolution of $p$-branes with $R |H| \ll 1$ in a flat FRW universe then Eqns.  (\ref{micvloop1a}) and (\ref{micvloop1b}) imply that the impact of the cosmology on brane dynamics is very small on a time scale $\lsim R$. 
Also since, for $H={\rm constant}$, the trajectory of a spherically symmetric $p$-brane is periodic, the effect of the cosmology averages to zero on each period of brane motion.  We will now assume that $H$ is no longer a constant and is given by $H(t)=H_0+\Delta H(t)$ where $H_0 > 0$ is 
a constant and $|\Delta H| < H_0$. For simplicity, in this section, we consider only the case of a circular cosmic string loop in 3+1 dimensions  ($N=3$, $p=1$, $i=0$). The generalization of our analysis for $N$, $p$ and $i$ arbitrary is straightforward. If we take $\Delta H=H_1(1-2 v^2(t+\theta))$, where $v$ is the microscopic velocity of the loop and $\theta$ is a constant phase, then Eqn.  (\ref{micvloop1b}) becomes
\be
\frac{dR}{dt} = R\left(1-2v^2(t)\right)\left(H_0 + H_1 \left(1-2v^2(t+\theta)\right)\right)\,.\label{micvloop1b1}
\ee
The evolution of the loop during one quasi period is hardly affected by the cosmological expansion and consequently the Minkowski space solution given by
\bq
r &=& r_0 \cos\left(\frac{t}{r^0}\right)\,,\\
v &=& -\sin \left(\frac{t}{r^0}\right)\,,
\eq
is still a very good approximation, in the $RH \ll 1$ limit, on time scales $\ll H^{-1}$.
Taking into account that $1- 2 v^2(t)= \cos(2t)$ and an averaging the r.h.s. of Eqn.  (\ref{micvloop1b1}) over one period one 
obtains the equation that describes the macroscopic evolution of circular loop
\be
\frac{d {\langle R \rangle}}{dt} =  \frac{H_1}{2}\cos (2 \theta) {\langle  R \rangle}\,, 
\ee
so that
\be
\langle R \rangle \propto  \exp \left(\alpha_2 H_1 t\right)\,, 
\ee
with $\alpha_2=\cos (2 \theta) /2$. We clearly see that the evolution of the universe may have an impact on the macroscopic 
evolution of cosmic string loops over cosmological time-scales, even if they are very small. To illustrate this we solved numerically the equations of motion for a cosmic string loop. In Fig. 3, we plot the results for the time evolution of the invariant radius of a loop with initial radius $R^0 H_0=0.002$ and for a Hubble parameter parametrized by by $H_0=2 H_1$ and $\theta=0$.  As expected this cosmology has indeed an impact on the evolution of the invariant radius, $R$, making its mean value increase after each period by the predicted amount. Of course, this is only valid in special cases, and in general such an effect is expected to be very small.


\section{\label{conc}Conclusions}

In this paper we studied the dynamics of special $p$-brane configurations in expanding and collapsing FRW universes with an arbitrary number of dimensions. We have shown that, in the case of spherically symmetric $p$-branes and a constant Hubble parameter, there are essentially two types of trajectories, closed periodic trajectories and open trajectories. We have obtained new analytical solutions for the trajectories in phase-space and computed the corresponding critical points as well as the root mean square velocity of the periodic $p$-brane solutions. We have shown that if $i \neq 0$ the solutions are no longer periodic 
with the $p$-brane radius evolving over cosmological timescales, even if $R|H| \ll 1$. 

We have also investigated the case where $H \neq {\rm constant}$ and we have found that for  $ -1 \le w < w_c$ with 
$w_c=(2-N)/N$ a critical radius may still be defined. In the $w \to w_c$ limit we found that $r_c H \propto (w_c-w)^{\gamma_c}$ with $\gamma_c=-1/2$. We also considered the special case of  a flat FRW universe dominated by a dense brane gas and 
determined the required conditions for accelerated expansion to take place. Finally, we discussed the impact that the large scale dynamics of the universe can have on the macroscopic evolution of very small loops showing that there are situations in which the evolution of the universe as a whole may affect the macroscopic $p$-brane dynamics over cosmological timescales.

\begin{acknowledgments}
This work is part of a collaboration between Departamento de F\'\i sica, Universidade Federal da Para\'\i ba, Brazil, and Departamento de F\'\i sica, Universidade do Porto, Portugal, supported by the CAPES-GRICES project. The authors also thank FCT,CNPQ and PRONEX-CNPq-FAPESQ for partial support. R.M. thanks CAPES for the fellowship BEX-0299/08-1. L.S. thanks 
FCT for the PhD grant SRFH/BD/41657/2007.
\end{acknowledgments}
\bibliography{pbranes}
\end{document}